# Coherence properties of light in highly multimoded nonlinear parabolic fibers under optical equilibrium conditions

Mahmoud A. Selim[1], Fan O. Wu[2], Georgios G. Pyrialakos[2], Mercedeh Khajavikhan[1], and Demetrios Christodoulides[1,*]

[1] Ming Hsieh Department of Electrical and Computer Engineering, University of Southern California, CA 90007, USA
[2] CREOL, The College of Optics and Photonics, University of Central Florida, Orlando, Florida 32816-2700, USA

*Corresponding author: demetri@creol.ucf.edu

**We study the coherence characteristics of light propagating in nonlinear graded-index multimode fibers after attaining optical thermal equilibrium conditions. The role of optical temperature on the spatial mutual coherence function and the associated correlation area is systematically investigated. In this respect, we show that the coherence properties of the field at the output of a multimode nonlinear fiber can be controlled through its optical thermodynamic properties.**

---

Multimode nonlinear fiber structures provide a promising testbed for observing a number of intriguing phenomena that have no counterpart in single-mode settings. These include, for instance, the beam self-cleaning effect [1–5] (Fig. 1(a)), novel Cherenkov radiation lines [6–8], and geometric parametric instabilities [9–11], to mention a few. What has initially motivated activities in this field is the promise for high bandwidth communication capabilities enabled by spatial mode-multiplexing schemes [12,13]. Naturally, this in turn incited a flurry of activities in the nonlinear domain in a quest for high-brightness sources [14]. Understanding how the optical field evolves in a nonlinear highly multimode system supporting perhaps hundreds or thousands of modes is by itself a challenging issue [15–19]. In general, to address this class of problems, numerical schemes can be deployed that could allow one to monitor the power evolution in each mode during propagation. Yet, at this point, such computational methodologies are not only time demanding but they also lend little physical insight, if any, as to underlying laws governing the behavior of these complex arrangements. Of interest will be to systematically comprehend the dynamics of such systems in an effort to exploit them to one's advantage. Quite recently, a self-consistent optical thermodynamic theory has been developed in order to predict and understand the convoluted behavior of these multimode settings [20–26]. Within this theoretical framework, one can then describe the collective macroscopic dynamics of such configurations using entropic principles that rely on statistical mechanics [20]. Importantly, once a nonlinear multimode optical structure reaches optical thermal equilibrium, its corresponding modal occupancies are governed by the celebrated Rayleigh-Jeans (RJ) distribution, which can be uniquely determined from initial conditions [21]. These formalisms can provide, in an effortless manner, the resulting modal occupancies at the output of any complex nonlinear multimode optical structure provided that the underlying nonlinearity is conservative. Moreover, the RJ distribution has been observed for the first time at the output of a multimode graded-index fiber in full accord with theory [27,28]. Fundamentally, one of the tenets of optical thermodynamics is ergodicity-an aspect resulting from the chaotic nature of nonlinear wave mixing processes that take place during evolution [20]. As such, at thermal equilibrium, the probability density function associated with the relative phases between optical modes is evenly distributed [20,22]. This, in turn, introduces complete decoherence among modes once thermalization ensues. At this point, one could naturally ask the following questions. At thermal equilibrium, how does the mutual coherence function vary across the core area of a multimode fiber? Are the resulting coherence properties of the output field related to the optical thermodynamic variables like temperature and chemical potential?

In this Letter, we investigate the degree of complex coherence function at the output of optically thermalized graded-index fibers. We show that the coherence function between two points in the near field output pattern can be drastically reduced as the optical temperature increases. The dependence of the coherence area is also studied as a function of temperatures and position within the core of the fiber, under both positive and negative temperature conditions. While the primarily focus of our study is on multimode graded-index nonlinear parabolic fibers, our results can be readily extended to other nonlinear multimode structures, like multi-core fiber systems and arrays as well as fibers having more general index profiles.

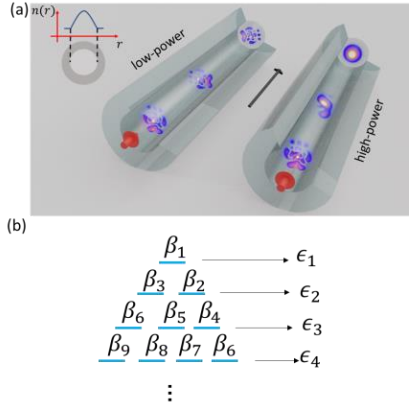

**Fig. 1.** (a) Optical thermalization. While in a linear fiber the modal power distributions remain constant during propagation, under weak nonlinear conditions the power is chaotically and ergodically reshuffled in a way that favors the lowest order modes when the temperature is positive. (b) Propagation constant diagram and respective degeneracy for a graded parabolic index multimode fiber.

We begin our analysis by considering how nonlinear beam propagation dynamics unfold in a weakly guiding parabolic optical fiber. Given that wave evolution in this system is paraxial, it can be effectively described by the following nonlinear Schrödinger equation:

$$i\frac{\partial \Psi}{\partial z} + \frac{1}{2k_0 n_1}\nabla^2 \Psi + k_0 n(x,y)\Psi + k_0 n_2 |\Psi|^2 \Psi = 0. \quad (1)$$

In Eq. (1), $\Psi(x,y,z)$ represents the slowly varying envelope associated with the optical field and $n(x,y) = n_1(1 - \Delta(x^2 + y^2)a^{-2})$ is the refractive parabolic index profile where $\Delta$ is the normalized index difference and $a$ is the fiber core radius and $n_1$ is the peak refractive index. Meanwhile $k_0 = 2\pi/\lambda_0$ stands for the free space propagation constant and $n_2$ denotes the Kerr nonlinear coefficient. This fiber supports in total $M$ modes whose eigenvalues $\epsilon_j$ (propagation constants) and modal profiles $|\psi_j\rangle$ can be determined from the eigenvalue equation $H_L|\psi_j\rangle = \beta_j|\psi_j\rangle$, where $H_L = (2k_0 n_1)^{-1}\nabla^2 + k_0 n(x,y)$ is the linear Hamiltonian of the waveguide system. Here, the eigenvalues are sorted in dec order according to $\beta_1 \geq \beta_2 \geq \beta_3 \ldots \geq \beta_M$, where $\beta_1$ and $\beta_M$ are the eigenvalues for the lowest-and highest-order optical modes, respectively. Note that this system exhibits two conserved quantities: the optical power $\mathcal{P} = \sum_{j=1}^{M}|c_j|^2$ and the longitudinal electromagnetic momentum flow (the so-called "optical energy") $U = -\sum_{j=1}^{M}\beta_j|c_j|^2$ [20–22], where $c_j(z)$ stands for the complex modal amplitude of the mode $|\psi_j\rangle$. These two invariants can be completely determined from the initial excitation conditions i.e., $\mathcal{P} = \sum_{j=1}^{M}|c_{j0}|^2$ and $U = -\sum_{j=1}^{M}\beta_j|c_{j0}|^2$ where $c_{j0} = \langle\psi_j|\Psi_o\rangle$, with $|\Psi_o\rangle$ representing the optical field at the input. Once $\mathcal{P}$ and $U$ are known, the optical temperature $T$ and chemical potential $\mu$ can be uniquely determined [21]. Note that the optical temperature is only related to the optical field and unrelated to the actual temperature of the fiber.

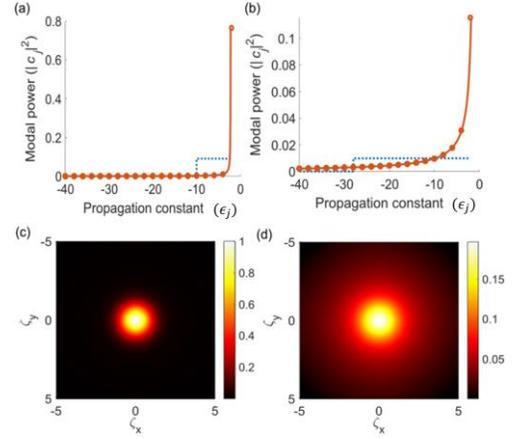

**Fig. 2.** (a) and (b) Resulting Rayleigh-Jeans distributions in a parabolic fiber at an optical temperature of $T = 0.021$ and $T = 0.085$, respectively. Spatial intensity distribution at the fiber output when (c) $T = 0.021$ and (d) $T = 0.085$. The blue dotted lines denote the projection of the excitation input in the linear modal basis of the fiber. The modes with propagation constants between (a) -10 and 0 and (b) -30 and 0 are equally excited. Here, $\mathcal{P} = 1$ and $M = 210$.

Subsequently, the thermalized modal distribution (expectation values) can be obtained through the RJ distribution, i.e., $|c_j|^2 = -T(\beta_j + \mu)^{-1}$. The modes in this parabolic fiber are here described within the orthogonal Hermite-Gauss set, that is:

$$\psi_{mn} = N_m N_h H_m(\zeta_x) H_h(\zeta_y) exp[-(\zeta_x^2 + \zeta_y^2)/2], \quad (2)$$

where $H_m(x)$ denotes a Hermite polynomial of order $m$ and $\zeta_x = \sqrt{V}x/a$ and $\zeta_y = \sqrt{V}y/a$ are normalized transverse coordinates. In Eq. (2), $V = k_0 a n_1 \sqrt{2\Delta}$ is the fiber V-number and $N_p = (a 2^p p!)^{-1/2}(V/\pi)^{1/4}$ is a normalization constant. In addition, the corresponding normalized propagation constants (eigenvalues) associated with this parabolic fiber are given by [29] $\beta_{mn} = -2(m + h + 1)$, where $m = 0,1,2,3,\ldots,Q$ and $h = 0,1,2,\ldots,Q - m$, where $Q = V/2$. In all cases, these indices satisfy $2(m + h + 1) \lesssim V$. From this latter equation, the propagation constants can be sorted in ascending order (see Fig. 1(b)), designated by a single eigenvalue level $\epsilon_j$. Each eigenvalue $\epsilon_j$ is associated with $j$ degenerate Hermite-Gauss eigenmodes in each polarization. Note that $\epsilon_j$ eigenvalues fall on an equidistant ladder-an aspect that greatly facilitates nonlinear four-wave mixing and thus optical thermalization. In this case, the total spatial intensity distribution emerging from a parabolic fiber can be represented as an incoherent superposition of all the modes, given that the power in each mode obeys the aforementioned RJ distribution: $|\Psi(\zeta_x, \zeta_y)|^2 = \sum_{j=1}^{M} -\frac{T}{\beta_j + \mu}|\psi_j(\zeta_x, \zeta_y)|^2$. Here, it is worth emphasizing that the term $-T/(\beta_j + \mu)$ is always positive and cannot be negative [21]. In the remaining of the paper, we will study the mutual coherence assuming that thermalization has already taken place and a statistical RJ distribution is established. Our investigation covers the complete temperature range, controlled by the initial excitation conditions, between $T = 0$, where all power occupies the fundamental mode as a condensate and $T = \infty$ where modal equipartition ensues. To investigate the coherence properties of a thermalized nonlinear multimode fiber, let us consider a GRIN parabolic fiber with a core radius of $50 \, \mu m$ and a numerical aperture of $NA = 0.2$, which supports in total $210$ modes at an operating wavelength of $\lambda_0 = 1550$ nm, without considering the polarization degeneracy. From here, the optical temperature and chemical potential can be uniquely determined [20,21] provided that the input excitation conditions are known. Note that throughout this paper we assume that $\mathcal{P} = 1$ and $M = 210$. As an example, Figs. 2(a) and 2(b) depict the thermalized modal power distribution for this particular parabolic fiber at an optical temperature of $T = 0.021$ and $T = 0.085$, respectively. Due to the nature of the RJ distribution, as the optical temperature increases the modal occupancies of the higher-order modes also increase. As a result, the corresponding spatial intensity profile at the output of a thermalized nonlinear parabolic fiber tends to broaden as the optical temperature increases (Figs. 2(c) and 2(d)). To describe the coherence properties of the resulting thermalized light within this optical fiber, we use the complex mutual coherence function $\mu_{12}$ [30,31]

$$\mu_{12} = \frac{\sum_{j=1}^{M} |c_j|^2 \psi_j(\zeta_x, \zeta_y) \psi_j^*(\zeta_x - \delta_x, \zeta_y - \delta_y)}{\sqrt{|\Psi(\zeta_x, \zeta_y)|^2 |\Psi(\zeta_x - \delta_x, \zeta_y - \delta_y)|^2}}, \quad (3)$$

where $|\Psi(\zeta_x, \zeta_y)|^2$ represents the intensity distribution of the optical beam at coordinates $(\zeta_x, \zeta_y)$, and $\delta_x$ and $\delta_y$ is the separation between two points under investigation along the $x$ and $y$ directions, respectively. Note that $\mu_{12}$ is a real function given that $\psi_j$ is also real under Hermitian conditions. As is well known, the magnitude of this function is always less than or equal to unity i.e., $|\mu_{12}| \leq 1$ with $|\mu_{12}| = 1$ representing total mutual coherence while for $\mu_{12} = 0$ total incoherence with respect to each other [30]. We begin by examining the mutual coherence function at a fixed coordinate point $(\zeta_x, \zeta_y)$ under positive temperature conditions. Figures 3(a)-3(f) illustrate the distribution of the coherence function at the output of a GRIN fiber having the same parameters as those considered in Fig. 2. As shown in Figs. 3(a)-3(c), at the center of the optical fiber $(\zeta_x = \zeta_y = 0)$ the mutual correlation function exhibits a circularly symmetric pattern, as one could anticipate. Nevertheless, as this figure shows, the coherence properties start to deteriorate with distance $(\delta_x, \delta_y)$ as the optical temperature increases. While at lower temperatures, this function is relatively broad, it quickly collapses at infinite temperatures $(T \to +\infty)$ where all modes convey the same amount of power, i.e., when equipartition of power takes place. In this regime, $\mu_{12}$ exhibits oscillations because of the strong presence of higher-order modes. Similar results are also presented in Figs. 3(d)-(f) for an off-center coordinate point $(\zeta_x, \zeta_y) = (1.25, 0)$. In this case, the mutual coherence function $\mu_{12}$ is now asymmetric with respect to $(\delta_x, \delta_y)$. Nevertheless, the same trend is observed as a function of temperature. We next consider the coherence properties between two points having a fixed separation $\sqrt{\delta_x^2 + \delta_y^2}$, while the coordinates $(\zeta_x, \zeta_y)$ are allowed to vary across the fiber cross-section. Figures 4(a)-4(d) illustrate how the $\mu_{12}$ function varies in this Figures 4(a) and 4(b) illustrate the mutual correlation for a pair of points separated by $(\delta_x, \delta_y) = (0.5, 0)$ at different optical temperatures. Evidently, as the temperature increases the coherence function around the center of a parabolic fiber decreases. Further away from the center of the fiber (i.e., $\zeta_x^2 + \zeta_y^2 \to \pm\infty$), the beam becomes almost fully correlated $\mu_{12} \approx 1$, since only the highest-order modes will contribute to the optical field. Furthermore, as the separation increases, the coherence function around the beam center will decrease, as shown in Figs 4(c) and 4(d). The coherence properties at the output of a thermalized optical fiber can also be investigated by means of a correlation area $A_c$ which in actual units can be defined according to [30,31] $A_c \equiv \rho_0^2 \int_{-\infty}^{\infty} \int_{-\infty}^{\infty} |\mu_{12}|^2 d\delta_x d\delta_y$, where $\rho_0 = a/\sqrt{V}$. This spatial coherence area can be regarded as the region over which the optical beam remains strongly coherent around a given point. On this basis, one can define an average coherence area

$$\overline{A_c} \equiv \frac{\int_{-\infty}^{\infty} \int_{-\infty}^{\infty} |\Psi(\zeta_x, \zeta_y)|^2 A_c(\zeta_x, \zeta_y) d\zeta_x d\zeta_y}{\int_{-\infty}^{\infty} \int_{-\infty}^{\infty} |\Psi(\zeta_x, \zeta_y)|^2 d\zeta_x d\zeta_y}. \quad (4)$$

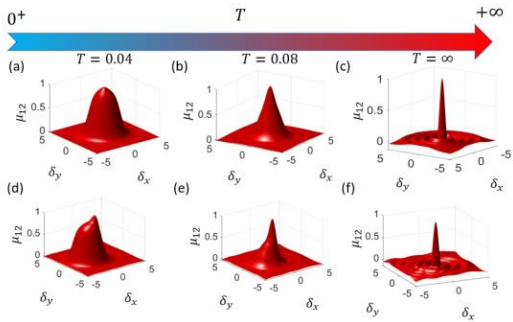

**Fig. 3.** Mutual coherence function around the center point $(\zeta_x, \zeta_y) = (0,0)$ at (a) $T = 0.04$ (b), $T = 0.08$ and (c) $T = \infty$. Figures 3(d)-3(f) depict similar results for the corresponding temperatures for an off-center point $(\zeta_x, \zeta_y) = (1.25, 0)$. Here, $\mathcal{P} = 1$ and $M = 210$.

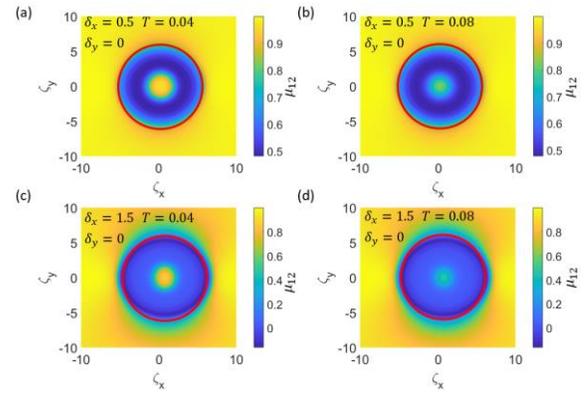

**Fig. 4.** Dependence of the mutual coherence function $\mu_{12}$ on the transverse spatial coordinates $\zeta_x$ and $\zeta_y$ when the separation distance is fixed. For each case, the separation and the corresponding temperature is labeled in the figure. The red circle indicates the core boundary of the GRIN optical fiber. Here, $\mathcal{P} = 1$ and $M = 210$.

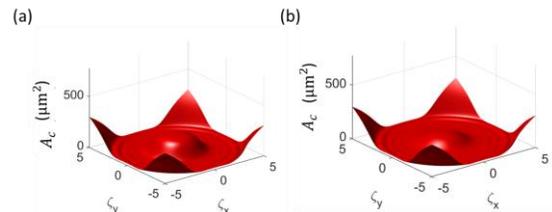

**Fig. 5.** Spatial coherence area $A_c$ at (a) $T = 0.03$ and (b) $T = 0.06$. The parameters used are the same as in Fig. 2.

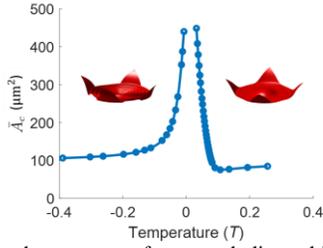

**Fig. 6.** Average coherence area for a parabolic multimode fiber as a function of the optical temperature. Here, $\mathcal{P} = 1$ and $M = 210$.

For the fiber considered in Fig. 2, Figs. 5(a) and 5(b) demonstrate that, in general, at the beam center the correlation area exhibits a local maximum. In addition, as the optical temperature increases the amplitude of this local maximum tends to decrease. On the other hand, as we move further away from the center of the fiber core, the coherence area again drastically increases because of the strong presence of higher-order modes. Interestingly, the average correlation area, obtained via Eq. (4), can be controlled by varying the optical temperature (or the initial excitation conditions), as shown in Fig. 6. Notably, as the absolute value of the optical temperature increases the effective coherence area decreases.

In conclusion, we have investigated the coherence properties of thermalized light emerging from a parabolic graded-index multimode optical fiber. In general, it was found that the coherence function between two points in the near-field output pattern is drastically reduced as the optical temperature increases. These results may provide new opportunities for controlling the coherence properties of thermalized light by means of optical thermodynamics.

**Funding.** This work was partially supported by the Office of Naval Research (ONR) (MURI: N00014-20-1-2789, N00014-18-1-2347, N00014-19-1-2052, N00014-20-1-2522), Air Force Office of Scientific Research (AFOSR) (MURI: FA9550-20-1-0322, MURI: FA9550-21-1-0202), National Science Foundation (NSF) (EECS-1711230, DMR-1420620, ECCS CBET 1805200, ECCS 2000538, ECCS 2011171), US Air Force Research Laboratory (AFRL) (FA86511820019), DARPA (D18AP00058), Army Research Office (W911NF-17-1-0481), Qatar National Research Fund (QNRF) (NPRP13S0121-200126), MPS Simons collaboration (Simons Grant No. 733682), W. M. Keck Foundation, US-Israel Binational Science Foundation (BSF: 2016381). G. G. Pyrialakos acknowledges the support of the Bodossaki Foundation.